\documentstyle[11pt,epsfig]{article}

\def\PL #1 #2 #3 {{\rm Phys. Lett.} {\bf#1} (#3) #2}
\def\NP #1 #2 #3 {{\rm Nucl. Phys.} {\bf#1} (#3) #2}
\def\ZP #1 #2 #3 {{\rm Z. Phys.} {\bf#1} (#3) #2}
\def\PRL #1 #2 #3 {{\rm Phys. Rev. Lett.} {\bf #1} (#3) #2}
\def\PR #1 #2 #3 {{\rm Phys. Rev.} {\bf#1} (#3) #2}
\def\MPL #1 #2 #3 {{\rm Mod. Phys. Lett.} {\bf#1} (#3) #2}
\def\RMP #1 #2 #3 {{\rm Rev.~Mod. Phys.} {\bf#1} (#3) #2}
\def\ifm{\ifmmode}

\def\als{\ifm \alpha_s \else $\alpha_s$\fi}
\def\go{\ifm \rightarrow \else $\rightarrow$\fi}

\def\eps{\ifm \epsilon \else $\epsilon $\fi}

\newcommand{\notp}{\ \hbox{{$p$}\kern-.43em\hbox{/}}}
\newcommand{\notE}{\ \hbox{{$E$}\kern-.43em\hbox{/}}}
\newcommand{\beq}{\begin{equation}}
\newcommand{\eeq}{\end{equation}}
\newcommand{\beqn}{\begin{eqnarray}}
\newcommand{\eeqn}{\end{eqnarray}}
\newcommand{\beqs}{\begin{eqnarray*}}
\newcommand{\eeqs}{\end{eqnarray*}}

\catcode`\@=11
\def\section{\@startsection{section}{1}{\z@}{3.5ex plus 1ex minus .2ex}
{2.3ex plus .2ex}{\large\bf}}
\def\thesection{\arabic{section}.}

\def\appendix{\setcounter{section}{0}
 \def\thesection{Appendix \Alph{section}:}
 \def\theequation{\Alph{section}.\arabic{equation}}}
\def\@citex[#1]#2{\if@filesw\immediate\write\@auxout{\string\citation{#2}}\fi
  \def\@citea{}\@cite{\@for\@citeb:=#2\do
    {\@citea\def\@citea{,\penalty\@m}\@ifundefined
       {b@\@citeb}{{\bf ?}\@warning
       {Citation `\@citeb' on page \thepage \space undefined}}%
\hbox{\csname b@\@citeb\endcsname}}}{#1}}

\def\citer{\@ifnextchar [{\@tempswatrue\@citexr}{\@tempswafalse\@citexr[]}}
\def\@citexr[#1]#2{\if@filesw\immediate\write\@auxout{\string\citation{#2}}\fi
  \def\@citea{}\@cite{\@for\@citeb:=#2\do
    {\@citea\def\@citea{--\penalty\@m}\@ifundefined
       {b@\@citeb}{{\bf ?}\@warning
       {Citation `\@citeb' on page \thepage \space undefined}}%
\hbox{\csname b@\@citeb\endcsname}}}{#1}}
\relax
 1
 1
 1

\begin{document}
\thispagestyle{empty}
\begin{flushright}
\hfill{CERN-TH/95-282}\\
\hfill{FERMILAB-Pub/95-369-T}
\end{flushright}
\vskip 2cm
\begin{center}
%{\large QCD Corrections to W Boson Plus Heavy Quark Production}
 QCD CORRECTIONS TO W BOSON PLUS HEAVY QUARK PRODUCTION
\vglue .4cm
%{\large at the Tevatron}
 AT THE TEVATRON
%{\ninerm (For 20\% Reduction to 8.5 $\times$ 6 in Trim Size)\\}
\vglue 1.4cm
\begin{sc}
Walter T. Giele, Stephane Keller\\
\vglue 0.2cm
\end{sc}
{\it Fermilab, MS 106\\
Batavia, IL 60510, USA}
\vglue 0.5cm
and
\vglue 0.5cm
\begin{sc}
 Eric Laenen\\
\vglue 0.2cm
\end{sc}
{\it CERN TH-Division\\
1211-CH, Geneva 23, Switzerland}
\end{center}

\vglue 1.5cm
\begin{abstract}
\par \vskip .1in \noindent
The next-to-leading order QCD corrections to
the production of a $W$-boson in association with a jet containing a heavy
quark are presented. The calculation is fully
differential in the final state particle momenta and includes
the mass of the heavy quark.  We study for the case of the Tevatron
the sensitivity of the cross section to the strange quark distribution
function, the dependence of the cross section on the heavy quark mass,
the transverse momentum distribution of the
jet containing the heavy quark, and the momentum distribution
of the heavy quark in the jet.
\end{abstract}

\vfill
\begin{flushleft}
CERN-TH/95-282\\
FERMILAB-Pub/95-369-T\\
November 1995
\end{flushleft}

\newpage
\setcounter{page}{1}
\section{Introduction}

\noindent The study of jet production in association with a vector boson
at hadron colliders has been succesful in the recent past.
The advantage of this signal over pure jet production is that
the lepton(s) from the vector boson decay can be used
as a trigger, such that jets can be studied free from jet-trigger bias.
Furthermore, the lower rate obviates the need for prescaling.
On the theory side, the recent progress in calculational
techniques to construct next-to-leading order (NLO) Monte-Carlo
programs~\cite{GGK93}
has allowed a meaningful confrontation with data~\cite{D0W}.

The tagging of heavy hadrons in the jet
offers a unique possibility of studying the hadronic structure inside the jet.
By considering jets where the {\em leading} hadron is tagged, a clear
connection
can be made with perturbative QCD.
At the parton level, the tagging of a heavy hadron corresponds to the tagging
of a heavy-flavor quark.  Experimentally,
the presence of a $D$ or $B$ meson is inferred through its decay products
(e.g. CDF has recently investigated $\gamma$ + D$^{*\pm}$ meson production,
where
the D$^{*\pm}$ meson was fully reconstructed~\cite{gamc}). Heavy flavor tagging
has become much
more efficient with the advent
of secondary vertex detectors~\cite{SVX90} and its importance
has been clearly demonstrated in the analysis that led to the top
quark discovery \cite{CDFD0}.  In the future, heavy flavor tagging will
continue to be an important analysis tool. It will provide more detailed
information about the event, as well as a test of the underlying QCD theory.
See, e.g., the recently outlined program to extract
parton densities exclusively from collider data \cite{wal95}.

If one demands the presence of a {\em charm} quark in the
jet recoiling against a $W$-boson, the signal is directly
sensitive to the strange quark distribution function in
the proton, at a scale of the order of
the $W$-mass.  A detailed investigation of this case has been performed in
\cite{BHKMR} with the shower Monte-Carlo program PYTHIA~\cite{PYT87}.

Replacing the charm quark by a {\em bottom} quark,
one could use this process as an alternative calibration of the $b$-quark
tagging efficiency.   This will be useful at luminosities achieved by
the Main Injector.
However, by far
the dominant contribution to $W$+bottom production is due to $W+b\bar{b}$
production, where the heavy quark pair is produced by gluon splitting.
Here the inclusion of the gluon to $B$ meson
fragmentation function is probably more important than the inclusion of
NLO effects to the $W$+bottom process.
This in turn would imply that this reaction could be used to constrain
this fragmentation function.

We present here the calculation of the QCD corrections up
to $O(\alpha_s^2)$ of the process $p\bar{p}\rightarrow W+Q$
where $Q$ is an heavy quark. We keep the mass of the heavy
quark explicit. In this letter we study the basic aspects of
this process. Full details of the calculation method and
more extensive phenomenological studies will be published
separately~\cite{GKL2}.

This letter is organized as follows.  In section 2, the method used to
calculate the QCD radiative corrections is briefly outlined and
some consequences of their inclusion are examined.
The impact of the higher order corrections on
the measurement of the strange quark distribution is discussed in section 3.
In section 4, we study several aspects of the behavior of the heavy quark
tagged jets.  Finally, in section 5 our conclusions are presented.

\section{Method}

\noindent
In this section the method used to calculate the
$O(\alpha_s^2)$ QCD corrections to
$p\bar{p}\rightarrow W +Q$ is outlined.
It consists of a generalization of the phase
space slicing method of Ref.~\cite{GG92}
 to include massive quarks.
One of the strengths of this method is that it allows for the implementation
of experimental cuts without the analytic recalculation of phase space
integrals.

\begin{figure}[hbt]
\vglue 2.8cm
\vbox{\includegraphics{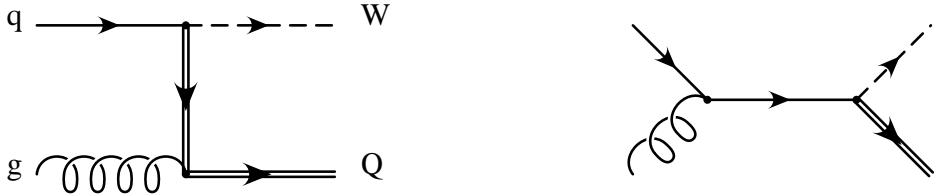} }
\caption{\it
The Born graphs for $W+Q$ production in $p\bar{p}$ collisions.}
\label{fig:Born_graphs}
\end{figure}

The leading order (LO) calculation is very simple and involves the two Feynman
diagrams
given in Fig.~\ref{fig:Born_graphs}.

The virtual corrections consist of
the interference between the lowest order diagrams and their one-loop
corrections. In order to regularize
the various singularities in the integrals over the loop momenta
we performed the integration in $d=4-2\epsilon$ dimensions.
The $d$-dimensional version \cite{Been} of the Passarino-Veltman reduction
formalism~\cite{PV} was used
to reduce tensor and vector integrals to scalar ones.
We made use of the algebraic
manipulation program FORM~\cite{FORM}
for much of the algebra.
Some of the scalar integrals were not yet available in the literature, they
will be listed in Ref.~\cite{GKL2}.
The ultraviolet singularities were absorbed through mass and coupling-constant
renormalization.
For the former we used the on-shell scheme
and for the latter the
$\overline{\rm MS}$ scheme modified such that the heavy quarks decouple
in the limit that small momenta flow into the heavy quark loops~\cite{CWZ}.
The remaining soft and collinear singularities,
appearing as $1/\eps^2$ and $1/\eps$ poles, factorize into a universal
factor multiplying the Born cross section.

The real corrections
consist of the contributions from all the subprocesses $i\,j \rightarrow
W\,Q\,k$,
where $i,j,k$ are massless partons, and the subprocess
$i\,j \rightarrow W\,Q\,\bar{Q}$~\cite{Mang}.
Some of these contributions exhibit soft and/or collinear singularities.
In this paper we treat the heavy quark as extrinsic to the nucleon, hence
we do not consider diagrams where the heavy quark is in the initial state.
The method we used to isolate the singularities consists of
slicing up the phase space and dividing it
into a hard region,
containing no singularities, and a region in which the final state parton
is either soft or emitted collinearly with one of the initial state partons.
Note that when a gluon is radiated
from the heavy quark, the collinear singularity is shielded
by the presence of the heavy quark mass.
The hard region is defined by the condition that all invariants
$s_{lm} = 2P_l\cdot P_m$, constructed out of the four-momenta
of any two neighboring partons/parton-heavy quark pairs
$l$ and $m$ in the
color-ordered subamplitudes~\cite{COLOR}, are larger than
a cut-off value $s_{\rm min}$.
The soft and collinear region corresponds to the case
where one or two of the $s_{ij}$ are smaller than $s_{\rm min}$.
In the hard phase space region, one
can work in four dimensions and perform the phase space integration
numerically.
In the soft and collinear region, the integration is done analytically in $d$
dimensions using soft and collinear approximations, which are valid in the
limit that $s_{\rm min}$ is small.  The cross section in this region again
factorizes into a universal factor multiplying the Born cross section.
The initial state collinear singularities are factorized
into parton distribution functions
in the ${\rm\overline{MS}}$ scheme, using the formalism
of crossing functions~\cite{GGK93}.

Note that the process $i\,j \rightarrow W\,Q\,\bar{Q}$ is quite
different from the other subprocesses: the heavy quark does not
originate from the $W$ vertex, and it is independent of
$s_{\rm min}$, because it is free from singularities.

Adding the real and virtual corrections leads to the cancellation
of all remaining singularities.  We checked gauge invariance for both the
virtual and real corrections.
One is finally left with a two-to-two particle contribution (consisting
of Born, soft-plus-collinear-plus-virtual, and
crossing function contributions) and
the two-to-three particle contribution in the hard region.

Before showing any numerical results,
we first list here the default choices we made for parameters and cuts in
producing the results of this paper.  Any deviation from these
choices will be indicated explicitly.
For the case of charm (bottom) we assumed three (four) light flavors
and no charm (bottom) quark distribution function.
We used both at LO and NLO the CTEQ3M~\cite{CTEQ3M} set of parton distribution
functions,
and the two-loop expression for the running coupling constant with
a four flavor, two-loop $\Lambda_{\rm QCD} = 0.239$ GeV, the
value supplied with the CTEQ3M set.
We implemented continuity across heavy flavor thresholds \cite{CT86}
using the parametrization of Ref.~\cite{ADMN}.
We used the Snowmass convention~\cite{SNOW} for the definition of a jet.
Our conditions on the transverse energy and pseudorapidity of the
jet were $E_{T}({\rm jet}) > 10$ GeV, $|\eta_{jet}| < 3$.
We took a jet cone size of $\Delta R=0.7$,
and implemented no cuts on the $W$.
We took the mass of the $W$-boson $m_W=80.23$~GeV,
the heavy quark mass $m$ equal to $1.7$~GeV for charm
and $5$~GeV for bottom. We used $V_{cs}=0.97$
and  $V_{cd}=0.22$ for the relevant Cabibbo-Kobayashi-Maskawa matrix elements.
We chose the factorization scale equal to the renormalization scale
and denote it by $\mu$, taking $\mu=m_W$.
At least one heavy quark was required to be inside of the jet, with the sign
of its electric charge correlated with the $W$ charge, as in the LO diagram.

\begin{figure}[hbt]
\vglue 7.5cm
\vbox{\includegraphics{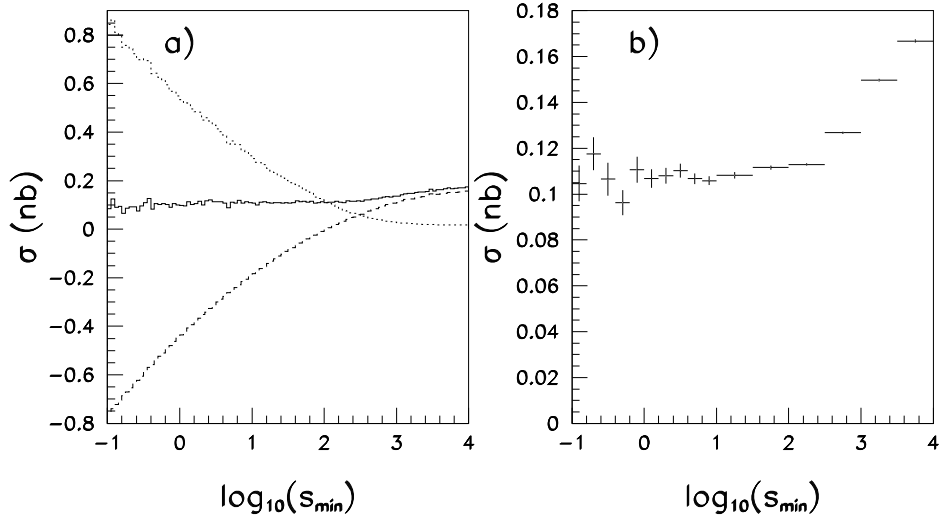}}
\caption{\it
$s_{\rm min}$-dependence of $p\bar{p}\rightarrow W^+$ + exclusive charm-tagged
one-jet
production. a) Solid histogram is for the
total cross section, dashed histogram is for its
two-to-two component,
and the dotted one is for its two-to-three
component.  b) Total cross section; the error bars represent the statistical
errors
from the Monte-Carlo integration.
}
\label{fig:smin}
\end{figure}
We first examine how the cross section changes when we vary the
arbitrary parameters $s_{\rm min}$ and $\mu$.
The dependence of the cross section and its two-to-two and two-to-three
components on the choice of $s_{\rm min}$ is shown in Fig.~\ref{fig:smin}a for
a wide $s_{\rm min}$ range.
It is clear from Fig.~\ref{fig:smin}a that each of the two components
depends strongly on the theoretical
cut-off $s_{\rm min}$, but at low $s_{\rm min}$ the cross section does not, see
also Fig.~\ref{fig:smin}b\footnote{Note that in this figure we used
for 0.1 GeV $<$  $s_{\rm min}$ $<$ 10 GeV
larger statistics and smaller bins than for $s_{\rm min}$ $>$ 10 GeV.}.
At high $s_{\rm min}$ the cross section varies because
the soft and collinear approximations used are no longer reliable.
We actually verified the $s_{\rm min}$ independence  in each
order in the expansion in $1/N_c$, where $N_c=3$ is the number of colors.
The results shown in the remainder of this paper are
averaged over $s_{\rm min}$ between 1 and 10 GeV.

\begin{figure}[hbt]
%\begin{center}
\vglue 7.5cm
\vbox{\includegraphics{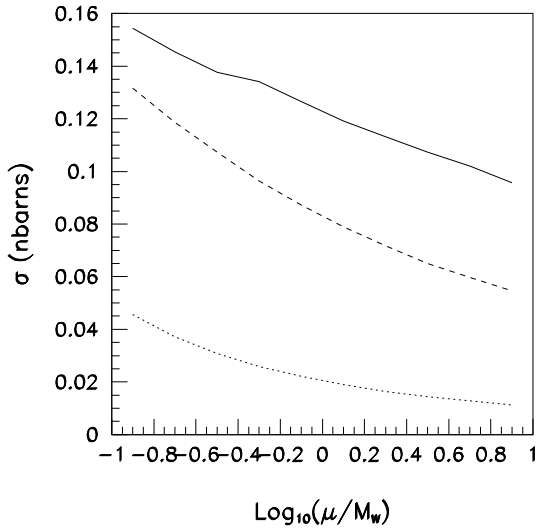}}
%\end{center}
\caption{\it
$\mu$-dependence for $p\bar{p}\rightarrow W^+$ + inclusive
charm-tagged one-jet production.
The solid line represents the NLO production
(minus the $W+c\bar{c}$ contribution), the dashed line the LO
production, and the dotted line the $W+c\bar{c}$ contribution.
}
\label{fig:mu}
\end{figure}
In Fig.~\ref{fig:mu} we show the renormalization/factorization
scale dependence of the
inclusive cross section, minus the $W+c\bar{c}$ contribution, along with
the LO contribution.  As expected, the inclusion of the NLO corrections
reduces
the scale sensitivity, albeit slightly.  We show the
$W+c\bar{c}$ contribution separately, because, as alluded to
earlier, this contribution is technically of leading order. It therefore
exhibits  a strong scale dependence, as can be seen in Fig.~\ref{fig:mu}.

\section{The Strange Quark Distribution in the Proton}

\noindent
In this section we discuss the effect that the inclusion of the NLO QCD
corrections
to $W$ + charm-tagged jet production has on constraining the strange quark
distribution
function $s(x,\mu)$ in the proton.  Here $x$ is the momentum fraction of the
strange quark in the proton, and $\mu$ is the factorization scale.
For the sake of clarity we briefly summarize
the study done in Ref.~\cite{BHKMR} using the shower
Monte-Carlo program PYTHIA.
At low scale $\mu$, the strange quark distribution function can
on the one hand
be inferred from the appropriate linear combination of
$F_2$ structure functions in neutrino
and muon deep inelastic scattering~\cite{CTEQ1M}.
On the other hand, it can also be determined from di-muon
events in neutrino deep-inelastic scattering (DIS)~\cite{CCF92}.
Using current experimental data sets, the two methods yield
a difference of about a factor of two for the strange quark distribution
function at low $\mu$.
It was suggested in \cite{BHKMR} that the strange quark distribution
function can also be constrained by determining the charm content of $W+1$~jet
events at the Tevatron, since the
leading order subprocess, $sg \rightarrow W c$, is
directly proportional to the strange quark distribution function.
In this measurement the strange quark
will effectively be probed at a larger scale $\mu \simeq M_W$.
At this higher scale, the difference between the two
strange quark distribution functions is smaller due to QCD evolution.
When relevant backgrounds are included and standard cuts are used,
the factor of two evolves into a difference of about 14\%
in the $W+c$ production cross section.
The charm tagging efficiency required to distinguish the two cases
at the one standard deviation level was found to be
about 10\% for 6000 $W+1$~jet events.

\begin{table}
\begin{center}
\begin{tabular}{|c|c|c|c|c|} \hline
  set           &mass (GeV)    & LO    & $WQ\bar{Q} $  &  NLO \\ \hline
  CTEQ1M        &$m_c$=1.7      & 96    &20             &161\\ \hline
  MRSD0'        &$m_c$=1.7      & 81    &20             &138\\ \hline
  CTEQ3M        &$m_c$=1.7      & 83    &20             &141\\ \hline
  CTEQ3M        & $m_b$=5.0     & 0.17   &9.09           &9.33 \\ \hline
\end{tabular}
\end{center}
\caption{\it
The $W$ + charm-tagged one-jet inclusive cross section in pb for
LO, $W+Q\bar{Q}$, and NLO (including the $W+Q\bar{Q}$ contribution)
using  different sets of
parton distribution functions.  The statistical uncertainty
from the Monte-Carlo integration is less than 1\%.
}
\end{table}

In Table 1 we give the NLO cross section for the parton distribution function
sets
CTEQ1M and MRSD0'.  The MRSD0' set derives its strange quark distribution
from the di-muon data, whereas the CTEQ1M set uses the DIS data.
Also shown is the result obtained with the more recent CTEQ3M set, which uses
the same assumption about the strange quark distribution as MRSD0'.
Comparing the CTEQ3M and MRSD0' sets, we see that the difference due to using
more
recent data sets in the global fit for the parton distribution functions is
small.
This is also reflected in the cross section for $W+ c\bar{c}$, which is
the same for all three sets.  We can conclude that the difference between
CTEQ1M and MRSD0' of 15.4 \% is due to the strange quark
distribution function.
This difference becomes 14.5\%  when one includes the
$W+b$ background (9 pb, almost all of it
coming from the gluon splitting contribution, see Table 1), and
assuming conservatively that each bottom quark
is mistagged as a charm quark.
This shows that the conclusions reached in Ref.~\cite{BHKMR}
are still valid at NLO. In both the NLO calculation and the PYTHIA analysis
about 50\% of the contributions are initiated by strange quarks.
One major difference is that PYTHIA suggests that the gluon splitting
contributes about 35\%,
whereas in the NLO calculation it is only about 15\%.
We found this number however to be quite
sensitive to the choice of factorization and renormalization scale
(recall we took $M_W$) in the gluon splitting contribution.
Further phenomenological study of
this question and others requires the inclusion of the $W$ leptonic
decay in our calculation.

\section{The Mass Dependence of Jet Production}

\begin{figure}[hbtp]
\vglue 7.5cm
\vbox{\includegraphics{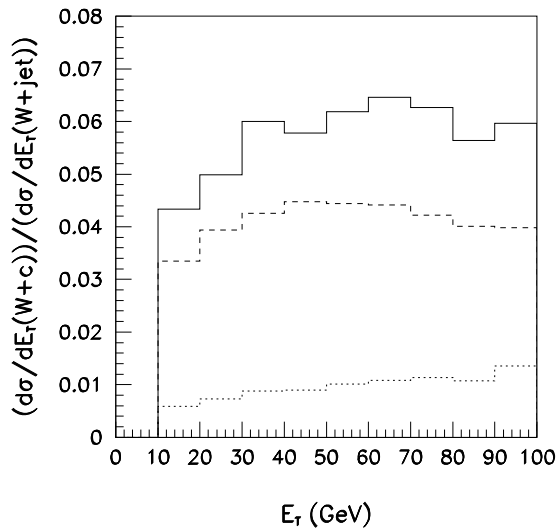}}
\caption{\it
Ratio of $W$ + charm-tagged inclusive one-jet
production over $W$ + inclusive untagged one-jet production
as function of the jet transverse energy.
The solid line is the NLO ratio and the dashed line the LO ratio.
The $W+c\bar{c}$ contribution to the NLO ratio is also shown
(dotted line).
}
\label{fig:ratio}
\end{figure}
\noindent
A comparison can be made between $W$ + untagged jet production\footnote{
For the untagged process we take five massless quark flavors.}  \cite{GGK93}
and $W$ + charm-tagged
jet production. The most obvious quantity to study in this
regard is the jet transverse energy ($E_T({\rm jet})$) distribution.
In Fig.~\ref{fig:ratio} we present the ratio of the charm-tagged jet
over the untagged jet $E_T({\rm jet})$-distribution for the LO and NLO cases.
At LO the charm-tagged jet is simply represented by a charm quark.
The ratio in Fig.~\ref{fig:ratio} has a characteristic shape which
can readily be understood at tree level.
At low $E_T({\rm jet})$ the charm-tagged jet rate is suppressed relative to
the untagged-jet rate due to its fermionic final state. The untagged-jet
rate is dominated by the gluonic final state which has a soft singularity,
that is absent for the fermionic final state.
At high $E_T({\rm jet})$ we again observe a relative suppression of the
charm-tagged jet because at LO this process has a gluon in the
initial state. At high $E_T({\rm jet})$
the dominant scattering in the untagged-jet rate
is due to quark-antiquark
collisions, again favoring the gluonic final state.
Apart from an approximate overall $K$-factor, the NLO cross section retains
these features although the identification with the LO parton model is lost.
Also shown in Fig.~\ref{fig:ratio} is the $W$ + $c\bar c$ contribution.
At low $E_T({\rm jet})$
its suppression is more pronounced
due to the charm quark pair production threshold.
At high $E_T({\rm jet})$ there is no suppression
for this process because it is instigated by a quark-antiquark
collision.

Thus we see that the $E_T({\rm jet})$ behaviour of tagged jets is basically
as expected.
We now turn to the mass dependence of the cross section.
In the LO calculation, one may take the charm mass to zero,
and, within Monte-Carlo errors, obtain a result which is identical to the
massive case.
However, we will now show that there are important mass effects
at NLO, especially when we look in detail at the tagged jets.
In fact, taking the heavy quark mass to zero
in the NLO calculation leads to a divergence.

\begin{figure}[hbtp]
\vglue 7.5cm
\vbox{\includegraphics{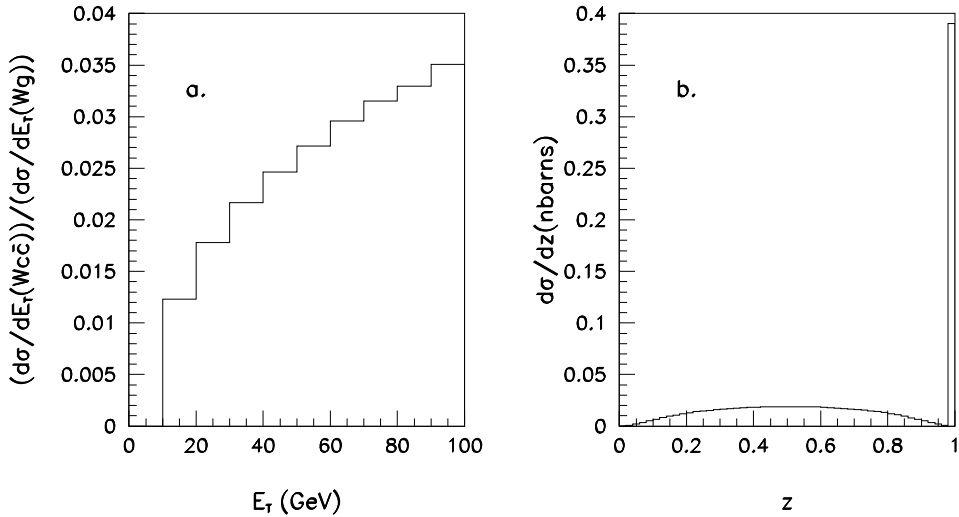}}
\caption{\it
a) Ratio of the $W+c \bar{c}$ component of the charm-tagged one-jet inclusive
cross section to the
$W+gluon$ cross section, as a function of the
jet tranverse energy.  b) The $z$-distribution of the
$W+c \bar{c}$ component.
}
\label{fig:wcc}
\end{figure}

Let us first consider the $W+Q\bar{Q}$ contribution.
Because the mass of the heavy quark regulates the collinear singularity,
it is expected that the strongest mass dependence will come from the collinear
region.
In this region the cross section factorizes
into the cross section
for $W + {\rm gluon}$ production  multiplied by a universal factor.
After integration over the invariant mass of the
heavy quark pair we find
that the mass dependent part of this universal factor has the following form:
\beq
\als \frac{N_c}{8\pi} P_{q\overline{q} \go g} (z)
\ln\left( \frac{M^2}{m^2} \right)
dz
\label{eq:lnm2}
\eeq
where $M$ is the upper limit of the
heavy quark pair invariant mass defining the collinear region,
and $P_{q\overline{q} \go g} (z)$ is the massless
Altarelli-Parisi~\cite{AP} splitting function:
\beq
P_{q\overline{q} \go g} (z) = \frac{2}{N_c} (z^2+(1-z)^2).
\label{eq:p}
\eeq
There is some ambiguity in
the definition of $z$.  Here, we choose the following:
\beq
z=\frac{E+P_{\parallel}}{E_{jet}+P_{jet}}
\eeq
where $E$ and $P_{\parallel}$ are the heavy quark energy and momentum
projected on the jet direction,
and $E_{jet}$ and $P_{jet}$ are the jet energy and momentum.
We have checked that other choices, such as $z=E_T(Q)/E_T({\rm jet})$,
do not change any of the conclusions in what follows.
In the strictly collinear limit $M$ is much smaller than the energy of the
gluon,
but in a leading logarithmic
approximation one may take $M$ to be of the order of $E_T({\rm jet})$.
The behavior of Eq.~(\ref{eq:lnm2}) can be seen
qualitatively in Fig.~\ref{fig:wcc}a, where the
ratio of the $W+c\bar{c}$ cross section over the $W+{\rm gluon}$ cross section
is
shown as a function of the transverse energy of the jet.
One can see an approximate logarithmic enhancement with increasing $E_T$,
as predicted by  the leading logarithmic approximation.
On the other hand, the $z$-distribution, plotted in Fig.~\ref{fig:wcc}b,
does not conform with
the $z$-dependence described by Eq.~(\ref{eq:lnm2}).
First, the peak at $z=1$ is due to events where the $\bar{Q}$ is not inside
of the jet, such that the whole jet is formed by the lone $Q$.
Second, the cross section is suppressed near $z=0$ and $z=1$ (excluding the
peak) due to terms in the collinear region that depend strongly on $z$,
but not on $M$.
\begin{figure}[hbtp]
\vglue 7.5cm
\vbox{\includegraphics{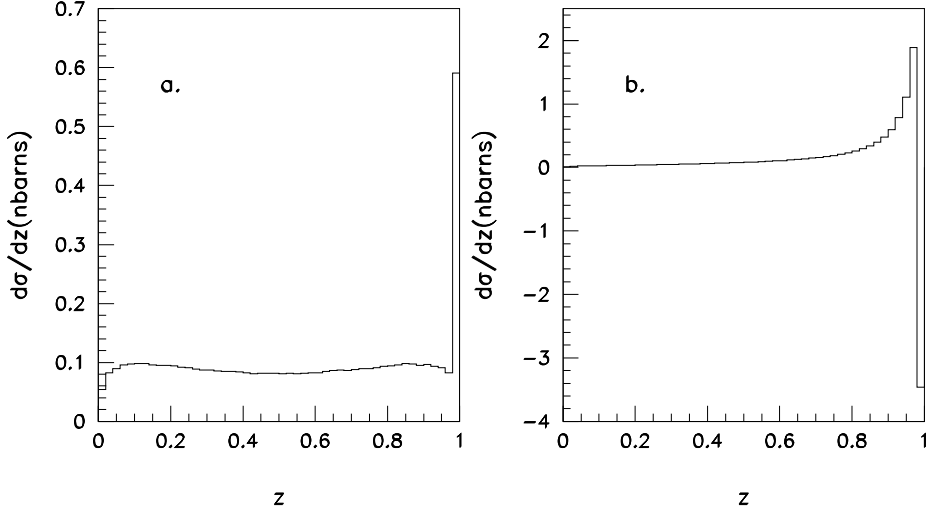}}
\caption{\it
The $z$-distribution of the charm-tagged one-jet inclusive cross section,
with $m=0.01$ GeV.  a) $Wc\bar{c}$ component. b) Total
contribution minus the $Wc\bar{c}$ component.
}
\label{fig:z}
\end{figure}
One way to enhance the effect of the
leading logarithm term of Eq.~(\ref{eq:lnm2})
in the $z$ distribution is to lower the mass of the heavy quark in our
calculation.
This is done in Fig.~\ref{fig:z}a, where we show the $z$-distribution
for the case $m=0.01$ GeV. Clearly it now resembles the functional form
of Eq.~(\ref{eq:p}) much closer.

The $\ln(m^2)$ term in Eq.~(\ref{eq:lnm2}) diverges in the limit of vanishing
quark mass, and is not cancelled by any other contribution~\footnote{
In the $W$ +1 jet calculation, this singularity
is cancelled by a companion collinear singularity in the quark-loop correction
to the outgoing gluon in the Born diagram. In our calculation this diagram is
not present.}.
In principle, any observable should be ``collinear safe'', i.e.
if the mass is taken to zero, the observable should
be finite and approach the massless result.
This is needed to describe situations where the relevant scale
is much larger than the heavy quark mass.
In the present case we are dealing with a final state divergence in $\ln(m^2)$
that should be factorized
into the fragmentation function of the gluon into a heavy hadron.
It is only after the proper introduction of the  fragmentation function in the
calculation that the
massless limit will be finite.
The evolution of fragmentation functions will
resum the large logarithms $\ln(E_T^2)$.
This problem is the final state version of the problem of heavy quark
distribution functions~\cite{ACOT94}.  It is beyond the scope of this short
letter, but we will discuss it in more detail in \cite{GKL2}.
Some studies in this regard were done in Ref.~\cite{Greco}.
Here, we simply keep the mass finite.

Let us now turn to the NLO single charm contribution (excluding the
$W+c\bar{c}$ contribution).  In the collinear region, the cross section again
factorizes, leading, after integration over
the invariant mass of the collinear partons, to the universal factor:
\beq
\als \frac{N_c}{8\pi} P_{qg \go q} (z)
\ln\left( \frac{M^2}{m^2} \right)
\label{eq:scll}
dz
\eeq
where $P_{qg \go q} (z)$ is the splitting function:
\beq
P_{qg \go q} (z) = \lim_{\delta\rightarrow 0}2 (1-\frac{1}{N_c^2})
\left(\left(\frac{1+z^2}{1-z}\right)\theta(1-z-\delta)
+(\frac{3}{2}+2\ln\delta)\,\delta(1-z)\right).
\label{eq:pqg}
\eeq
In Fig.~6b we show for the single charm quark contribution the
$z$-distribution, with the contribution of Eq.~(\ref{eq:scll}) enhanced
by taking  $m = 0.01$ GeV. Note that it resembles the functional
form in Eq.~(\ref{eq:pqg}).  From Eq.~(\ref{eq:pqg}) one derives
\beq
\int_0^1 P_{qg \go q} (z) dz= 0 \ ,
\label{eq:qsum}
\eeq
which must hold for the probability to find a quark in a quark of the same
flavor to be one \cite{AP}.
{}From Eqs.~(\ref{eq:qsum}) and (\ref{eq:scll})
we can now make the important observation that as long as the cuts
on the heavy quark-tagged jet are such that all $z$-values are allowed to
contribute,
there are no large logarithms $\ln(E_T^2/m^2)$.
\begin{figure}[hbtp]
\vglue 7.5cm
\vbox{\includegraphics{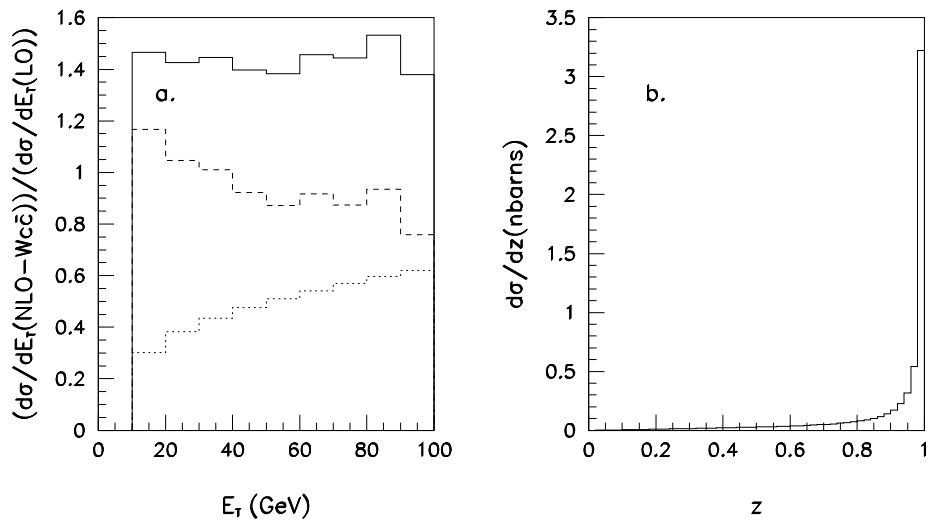}}
\caption{\it
a) $E_T$ distribution ratio of NLO charm-tagged one-jet inclusive
cross section to LO one.
Solid line has no restrictions on the $z$-integration,
the dotted and dashed line have restrictions of $z<0.9$ and $z>0.9$
respectively.
b) $z$-distribution of NLO charm-tagged one-jet inclusive cross section.
The $W+c\bar{c}$ component is not included in these plots.
}
\label{fig:z+et}
\end{figure}
However, if the cuts are such that the $z$-integration is restricted, or
convoluted with a $z$-dependent function, some
$\ln(E_T^2/m^2)$ terms will remain.  An example of such a
convolution is the $E_T$ distribution of the heavy quark itself.
All this is illustrated in Fig.~\ref{fig:z+et}a,
where two cases of $z$-restrictions ($z > 0.9$ and $z < 0.9$) are plotted
in addition to the all-$z$ case.
Note that for the all-$z$ case
the ratio is indeed essentially independent of $E_T({\rm jet})$, but that for
the $z$-restricted cases the logarithmic dependence is apparent.
For completeness we show in Fig.~\ref{fig:z+et}b the $z$-distribution
for the single charm contribution for $m=1.7$ GeV.

Thus care must be taken when calculating tagged cross sections
in determining whether or not there are large logarithms present
due to restrictions on the $z$-integration.
Such restrictions would also necessitate the introduction of the appropriate
fragmentation function to absorb the $\ln(m^2)$ terms, as in the $WQ\bar{Q}$
case.

Note that for the example of the jet transverse energy distribution
given in Fig.~4 there is no constraint on the $z$-integration, so that the
only logarithmic mass term is due to the $Wc\bar{c}$ contribution.

\section{Conclusions}

\noindent
We have completed the first calculation
of the QCD corrections to $O(\alpha_s^2)$ of the reaction
$p\bar{p}\rightarrow W +Q$.  In this short paper we briefly
summarized the method of calculation.
We demonstrated that the inclusion of the NLO corrections
does not change the conclusions of Ref.~\cite{BHKMR} about constraining
the strange quark distribution function using
$W$ + charm-tagged jet events at the Tevatron.
However, since we now have a NLO calculation, this procedure will be able to
constrain the NLO strange quark distribution, once a reasonable data
sample is collected.
Finally we studied the $E_T$ distribution
of the jet containing the heavy quark and the mass dependence of the cross
section.  We noted the need to include the
leptonic decay of the $W$ and the heavy hadron fragmentation functions
in our calculation, in order to be able to do more extensive phenomenological
studies.


\begin{thebibliography}{9}

\bibitem{GGK93}\label{GGK93}
W.T.~Giele, E.W.N.~Glover, and D.A.~Kosower, \NP B403 633 1993 .

\bibitem{D0W}\label{D0W}
S.~Abachi et al (D0 collaboration), \PRL 75 3226 1995 .

\bibitem{gamc}\label{gamc}
R.~Blair for the CDF collaboration, FERMILAB-CONF-95/245-E.
Presented at 10th
Topical Workshop on Proton-Antiproton Collider Physics,
Batavia, IL, 9-13 May 1995.

\bibitem{SVX90}
F.~Abe et al (CDF Collaboration), Nucl. Instrum. Meth.
{\bf A289} (1990) 388; \NP B27 246 1992 \ (Proc. Suppl).

\bibitem{CDFD0}\label{CDFD0}
F.~Abe et al (CDF collaboration), \PRL 74 2626 1995 ;
S.~Abachi et al (D0 collaboration), \PRL 74 2632 1995 .

\bibitem{wal95}\label{wal95}
W.T.~Giele. and E.W.N.~Glover, FERMILAB-CONF-95-168-T.
Presented at 10th
Topical Workshop on Proton-Antiproton Collider Physics,
Batavia, IL, May 1995.

\bibitem{BHKMR}\label{BHKMR}
U.~Baur, F.~Halzen, S.~Keller, M.L.~Mangano and K.~Riesselmann,
\PL  B318 544 1993 .

\bibitem{PYT87}\label{PYT87}
H.--U.~Bengtsson and T.~Sj\"ostrand, Comp. Phys. Comm. {\bf
46} (1987) 43; T.~Sj\"ostrand, preprint CERN-TH.6488/92.

\bibitem{GKL2}\label{GKL2} W.T. Giele, S. Keller and E. Laenen,
in preparation.

\bibitem{GG92}\label{GG92}
W.T.~Giele and E.W.N.~Glover, \PR D46 1980 1992 .  For equivalent methods
see H.~Baer, J.~Ohnemus, and J.~F.~Owens, \PR D40 2844 1989 ;
S.~Frixione, M.L.~Mangano, P.~Nason and G.~Ridolfi, \NP B403 633 1993 .

\bibitem{Been}\label{Been} W.~Beenakker, Ph.D. thesis, unpublished.

\bibitem{PV}\label{PV} G.~Passarino and M.~Veltman, \NP B160 151 1979 .

\bibitem{FORM}\label{FORM}
FORM2 by J.A.M.~Vermaseren, Published by
Computer Algebra Netherlands (CAN), Kruislaan 413,
1098 SJ Amsterdam, The Netherlands.

\bibitem{CWZ}\label{CWZ} J.C.~Collins, F.~Wilczek, and A.~Zee, \PR D18 242 1978
;

\bibitem{Mang}\label{Mang} M.L.~Mangano, \NP B403 536 1993 .

\bibitem{COLOR}\label{COLOR}
F.A.~Berends and W.T. Giele, \NP B306 759 1988 ;
M.L.~Mangano, \NP B309 461 1988 .

\bibitem{CTEQ3M}\label{CTEQ3M}
H.L.~Lai et al,  \PR D51 4763 1995 .

\bibitem{CT86}\label{CT86} J.C.~Collins and W-K.~Tung \NP B278 934 1986 .

\bibitem{ADMN}\label{ADMN} G.~Altarelli, M.~Diemoz, G.~Martinelli and
P.~Nason, \NP B308 724 1988 .

\bibitem{SNOW}\label{SNOW}
J.E.~Huth et al, in proceedings of the Snowmass Workshop,
{\sl High Energy Physics in the 1990's}, Snowmass, Colorado,
July 1990, p. 134.

\bibitem{CTEQ1M}\label{CTEQ1M}
J.~Botts et al, \PL 304B 159 1993 .

\bibitem{CCF92}\label{CCF92}
W.H.~Smith et al (CCFR Collaboration), in proceedings
of ICHEP 92, Dallas, 1992, J.R. Sanford, ed.

\bibitem{AP}\label{AP} G.~Altarelli and G.~Parisi, \NP B126 298 1977 .

\bibitem{ACOT94}\label{ACOT94}
M.A.G.~Aivazis, J.C.~Collins, F.I.~Olness and W-K.~Tung,
\PR D50 3102 1994 .

\bibitem{Greco}\label{Greco}
M.~Cacciari and M.~Greco, \NP B421 530 1994 ;
B.~Kniehl, M.~Kr\"{a}mer, G.~Kramer and M.~Spira, \PL 356B 539 1995 ;
M.~Cacciari and M.~Greco preprint DESY 95-103.
\end{thebibliography}
\end{document}